\documentclass{iau}

\usepackage{amsmath}
\usepackage{graphicx}
\usepackage{multirow}

\usepackage{amsmath}
\usepackage{multirow}
\usepackage{txfonts}

\def\HI{{\tt HI}}

\usepackage{printlen}
\usepackage{layouts}

\newcommand{\kms}{$\,$km~s$^{-1}$}

\newcommand{\msun}{{${\rm M}_\odot$}}

\newcommand{\cmsq}{cm$^{-2}$}

\newcommand{\eg}{\mbox{e.g.}}

\newcommand{\ngc}{\mbox NGC~1427A}

\begin{document}

\lefttitle{The MeerKAT Fornax Survey}
\righttitle{removal of \HI\ from galaxies in the Fornax cluster}

\jnlPage{1}{7}
\jnlDoiYr{2021}
\doival{10.1017/xxxxx}

\aopheadtitle{Proceedings IAU Symposium}
\editors{D.~J. Pisano,  J. Healy \&  S. Blyth, eds.}

\title{The MeerKAT Fornax Survey: removal of \HI\ gas from galaxies in the  Fornax cluster}

\author{F. M. Maccagni, P. Serra}
\affiliation{INAF -- Osservatorio Astronomico di Cagliari, via della Scienza 5, 09047, Selargius (CA), Italy}

\begin{abstract}
The MeerKAT Fornax Survey is  conducting a thorough examinationof the nearby Fornax galaxy cluster to understand how galaxies lose their cold gas and stop forming stars in low-mass clusters (M$_{\rm vir} \leq 10^{14}$ M$_\odot$). We are doing so through very deep (down to $\sim 10^{18}$ cm$^{-2}$) and high resolution (up to$\sim~ 1$ kpc and 1 km s$^{-1}$) MeerKAT observations of neutral atomic hydrogen gas (HI) in a $1\times2$ Mpc$^2$ region centred on Fornax.
At the time of writing, the survey is $88\%$ complete. Initial analysis of the cluster's central area has unveiled the widespread existence of previously unseen \HI tails and clouds. Some of the \HI\ is clearly being removed from Fornax galaxies as they interact with one another or with the intra-cluster medium. We present a sample of galaxies with long, one-sided, star-less \HI\ tails (of which only one was previously known) radially oriented within the cluster. The properties of these tails represent the first conclusive evidence that ram pressure is a key force shaping the distribution of \HI\ in the Fornax cluster. Furthermore, interactions within the Fornax environment shape the \HI\ mass function, the \HI\ content of dwarf galaxies and determines how the \HI\ sustains the nuclear activity of some cluster members and neighbouring galaxies.

\end{abstract}

\begin{keywords}
galaxies: clusters: individual: Fornax, galaxies: evolution, galaxies: interactions, galaxies: ISM
\end{keywords}

\maketitle

\section{Introduction}

Many galaxies in the Universe live in groups and galaxy clusters. There several phenomena affect their evolution, galaxyies may collide or merge, but they can also interact hydro-dynamically with the gaseous medium that fills the environment, given that the medium has sufficient density to exert a drag force on the galaxies that move within it.

The morphology and gas content of galaxies are strongly affected by the properties of the surrounding environment, such as the number density of galaxies and their motions relative to one another and in the more or less dense intergalactic medium (IGM). Nevertheless, it is unclear which physical phenomena strip away the gas from galaxies and which regulate its accretion onto them and over which timescales. For instance, do tidal interactions or ram pressure stripping play a dominant role in gas removal? How to these processes vary with galaxy mass, trajectory and location within the clusters and groups?

Neutral atomic hydrogen (\HI) has long served as a prime tracer of environment-driven galaxy evolution(\eg Gavazzi 1989, Kenney et al. 2004).  A crucial property of \HI\ is that it can be detected all the way out to the outskirts of galaxies, at the boundary between the galactic bodies and the environment. There, reaching low-column \HI\ densities ($\leq 10^{20}$~\cmsq) with kpc-resolution, is possible to study in detail on-going episodes of gas removal as well as accretion~(\eg\ Sancisi et al. 2008) and link them to the environmental processes driving them. \HI\ is also the gas phase from which molecular gas forms, and is therefore a key ingredient of the baryon cycle regulating the star formation of galaxies~(\eg\ Walter et al. 2020).

\begin{figure}[t]
  \centerline{\vbox to 1pc{\hbox to 1pc{}}}
  \centering
  \includegraphics[width=\textwidth]{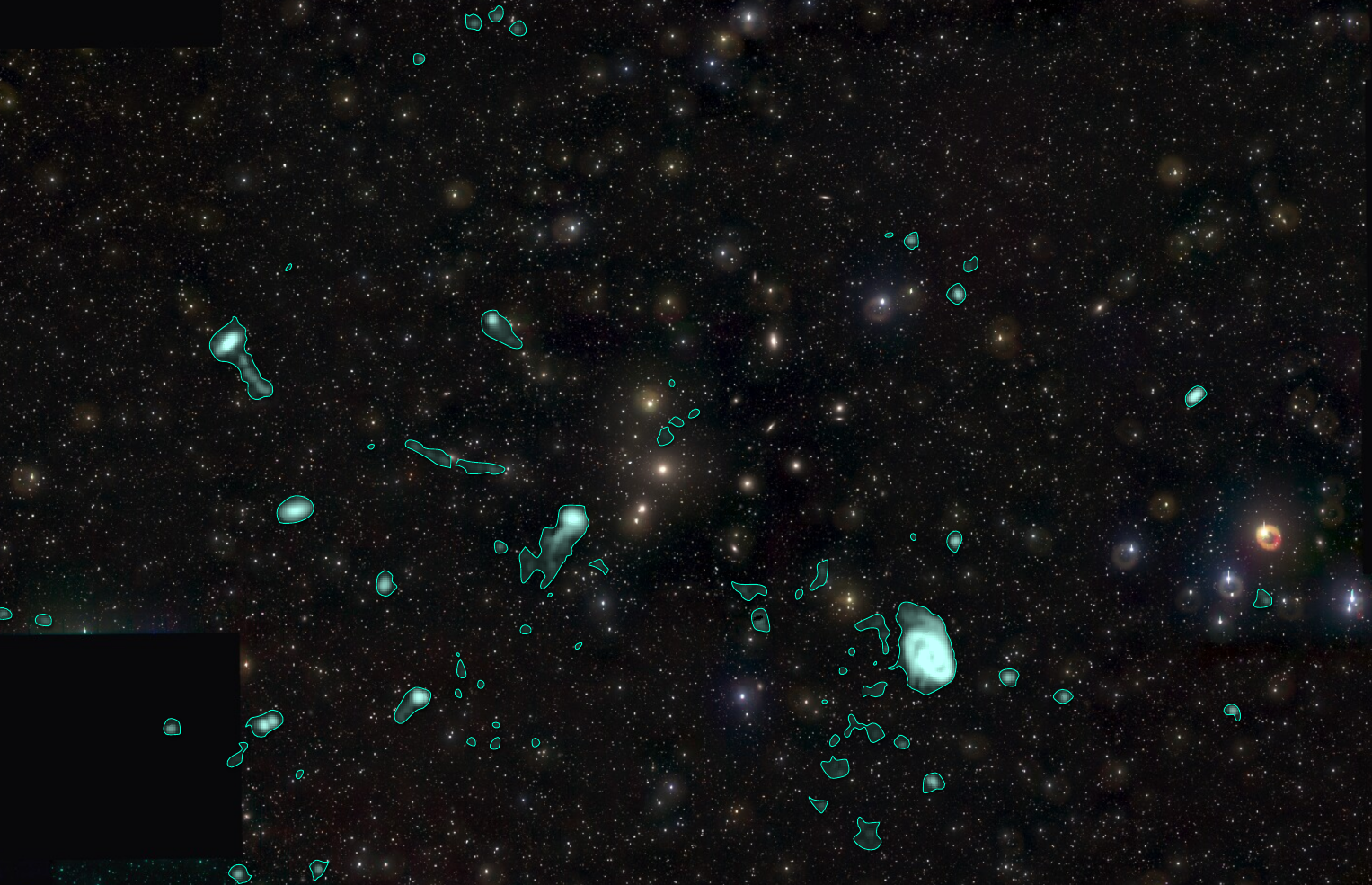}
  \caption{{\em ugri} image of the Fornax Cluster from the Fornax Deep Survey, overlaid with \HI\ emission contours ($1\times 10^{18}$~\cmsq at 66'') from the MeerKAT Fornax Survey.}
  \label{fig1}
\end{figure}

The Fornax cluster (Fig.~\ref{fig1}) is an ideal target for studying environmental processes through \HI\ observations because of its proximity (20 Mpc, Jensen et al. 2001), size and evolutionary state. The cluster is not relaxed with the massive group of NGC 1316 ($1.6\times 10^{13}$\msun) falling in from the southwest, indicating that, it remains in an active stage of formation. In the low-mass regime of the Fornax cluster $M_{vir} = 5 \times 10^{13}$~\msun~(Drinkwater et al. 2001), it is still unclear to what extent, through which processes and over what timescales galaxies evolve because of environmental interactions. For this reason we are undertaking the MeerKAT Fornax Survey Survey\footnote{\url{https://sites.google.com/inaf.it/meerkatfornaxsurvey}} which is designed to map, in unprecedented detail, the distribution and kinematics of \HI\ within and between galaxies in the Fornax cluster (Serra et al. 2023). 


\section{The MeerKAT Fornax Survey}

The goal of the MeerKAT Fornax Survey (MFS) is to determine whether galaxies undergo significant interactions and undergo significant environment-driven evolution and, if so, to estabrlish the timescales and physical processes involved. How do galaxies lose their cold gas? why do they stop accreting fresh gas? how do these processes impact their evolution? In addition to these primary objectives, our MeerKAT observations also enable complementary studies of star formation, AGNs and magnetic fields within galaxies and the intra-cluster medium of Fornax. 

\begin{figure}[t]
  \centerline{\vbox to 1pc{\hbox to 1pc{}}}
  \includegraphics[width=\textwidth]{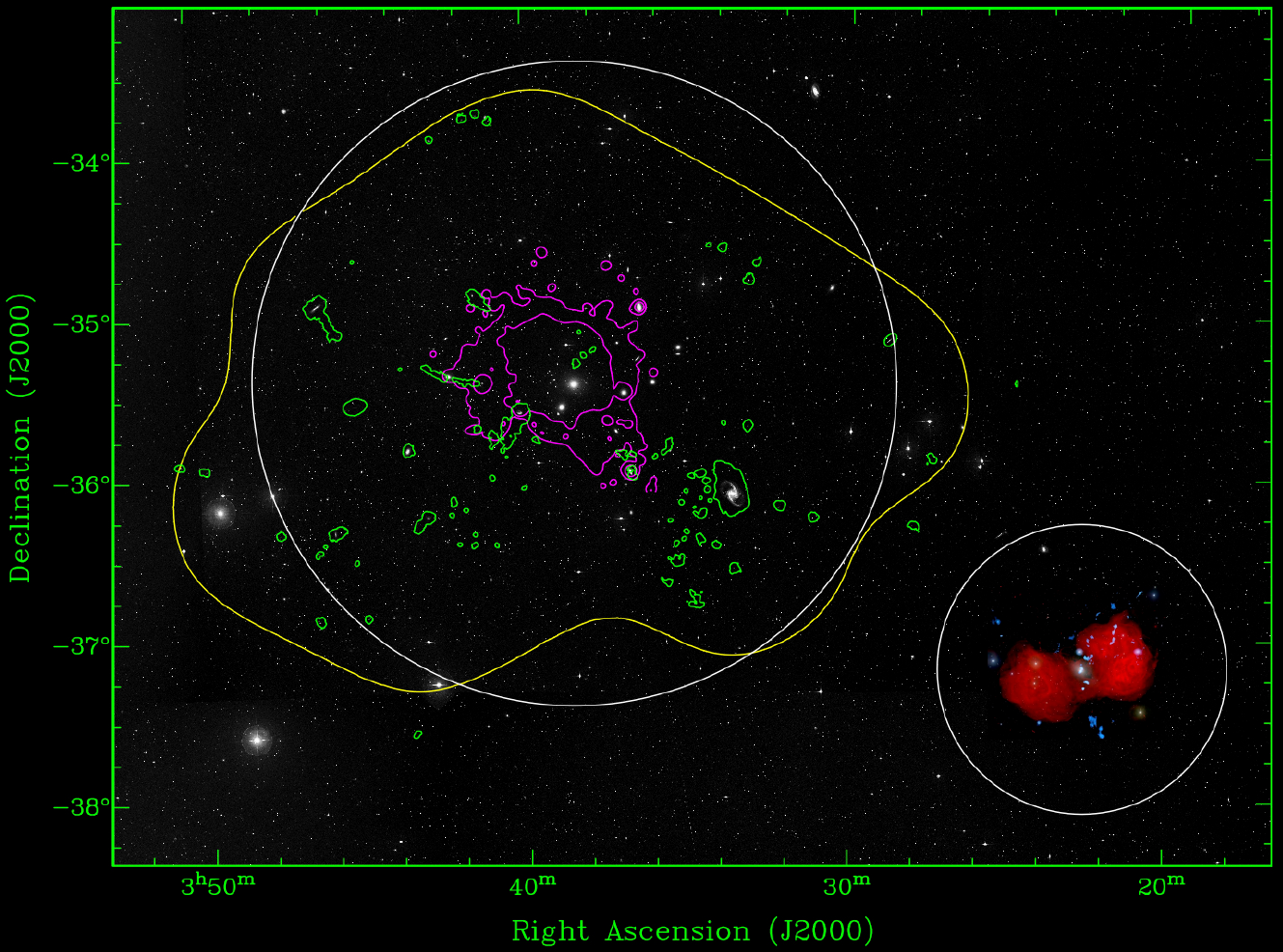}
  \centering
  \caption{The Fornax cluster with the footprint of the MeerKAT Fornax Survey observations reduced so far overlaid (yellow contours). \HI\ detections are shown in green ($1\times 10^{18}$~\cmsq at 66''). The grey circles mark the 700 kpc virial radius of the cluster and the 380 kpc virial radius of the infalling NGC 1316 group. The X-ray emission associated with the cluster is shown in magenta.}
  \label{fig:survey}
\end{figure}

The MFS consists of L-band \HI\ and full-Stokes radio continuum MeerKAT~(Mauch et al. 2020) observations of a $\sim 12$ deg$^2$ area centered on the Fornax galaxy cluster and extending in the southwest toward the NGC 1316 group, which hosts the bright and extended radio AGN Fornax A (Fig.~\ref{fig:survey}). The MFS improves sensitivity and resolution over previous \HI\ observations~(Loni et al. 2020) by one to two orders of magnitude, providing a complete census of ongoing environmental interactions in Fornax by delivering \HI\ cubes, images, velocity fields, and velocity dispersion maps with an angular resolution from $\sim 10$'' ($\sim 1$ kpc) to $\sim 100$'' ($\sim 10$ kpc), a velocity resolution of $\sim 1.4$~\kms, and a column density sensitivity between $\sim 8\times 10^{17}$ and $\sim 5\times  10^{19}$\cmsq\ depending on the angular resolution. The survey is sensitive to \HI\ detections down to M$_{\rm HI}\sim6 \times 10^5$\msun\ across the entire cluster. At the time of writing, 80 of the $91$ $\sim1$ deg$^2$ fields have been observed, making the survey $88\%$ complete. The observations reduced so far (yellow contour in Fig.\ref{fig:survey}) cover the full virial radius of the cluster, while observations of the NGC 1316's are on-going. Part of this group has been observed with a single pointing at low spectral resolution (44~\kms) during MeerKAT's commissioning. Different results obtained combining the \HI\ and $1.4$GHz continuum observations are summarised in Sect.~\ref{fornaxa}. For a detailed description of the data reduction of the spectral line observations and of the first data release see~(Serra et al. 2023). 

The MFS is completely changing our picture of Fornax. Until recently, Fornax appeared to contain very little \HI, most of which was confined to the stellar body of a few galaxies at the cluster outskirts and showed limited evidence of ongoing interactions. In contrast, our MeerKAT data reveal a dynamic environment showing several \HI\ tails and clouds associated to the cluster galaxies and tracing different environmental interactions which have a strong impact on their evolution.

The MFS so far focused on the study of the morphology and kinematics of the \HI\ detections, and of the dwarf galaxies of the cluster. For the first time, it has been shown that, in a small cluster like Fornax: i) the distribution and kinematics of \HI\ in gas-rich can be affected galaxies can be affected by ram pressure once they have already been perturbed by a tidal interaction~(Serra et al. 2023); ii) dwarf galaxies lose their \HI\ rapidly~(a few hundreds million years, Kleiner et al. 2023).



\section{Tidal effects and ram pressure rapidly shaping the galaxies in the cluster}

\begin{figure}[t]
  \centerline{\vbox to 1pc{\hbox to 1pc{}}}
  \includegraphics[width=\textwidth]{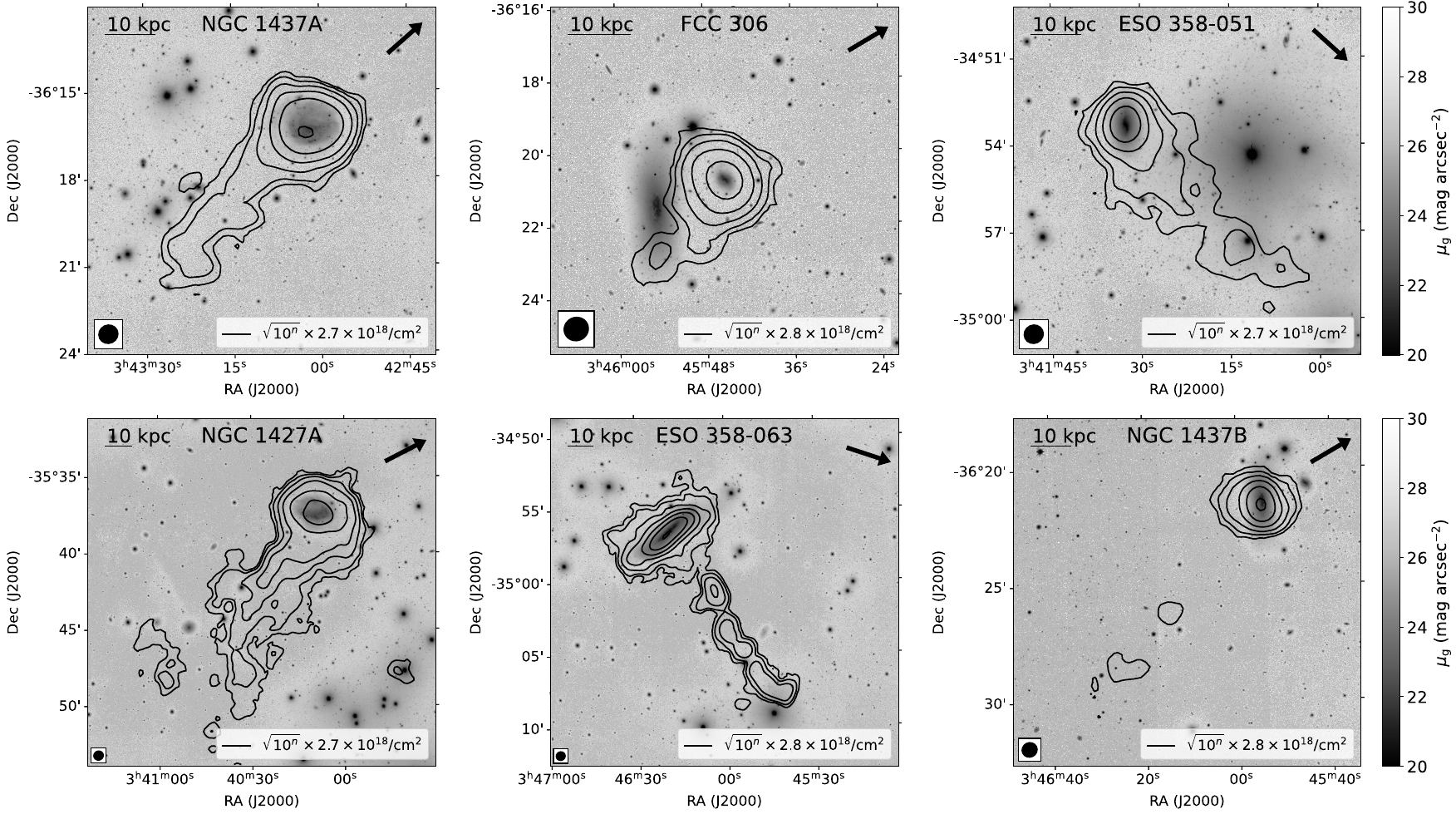}
    \centering
  \caption{MeerKAT \HI\ contours overlaid on a g-band image from the FDS for six galaxies with a one-sided \HI\ tail. The galaxy name is shown at the top of each panel. \HI\ contour levels are indicated in the bottom right. The black ellipse in the bottom left represents the $41$'' PSF. The arrow in the top right points toward the center of the cluster~(Serra et al. 2023).}
  \label{fig:ensemble}
\end{figure}

Figure~\ref{fig:ensemble} shows the \HI\ contours at a resolution of $41$'' ($\sim 4$ kpc at the distance of Fornax) overlaid on a g-band optical image taken from the Fornax Deep Survey~(FDS, Iodice et al. 2016) of six galaxies with extended ($\geq 20$ kpc) \HI\ tails, all directed radially within the cluster. The overall analysis of the galaxies' stellar and \HI\ morphologies and kinematics indicates that all galaxies with an \HI\ tail are tidally disturbed, suggesting that this is a necessary condition for generating the tail in the first place. Then, ram-pressure acts on the tails stretching them to their observed length~(Serra et al. 2023).

To accurately understand the delicate balance and interplay between these tidal interactions and ram pressure, it is necessary to study these galaxies in detail, individually. Here, we focus on \ngc\ (bottom left panel of Fig.~\ref{fig:ensemble}), which which exhibits a spectacularly complex \HI\ morphology, with a warped disk and an extended (70 kpc) \HI\ tail in the South-East of its disk.
 
\subsection{The case of \ngc}

\begin{figure}[t]
  \centerline{\vbox to 1pc{\hbox to 1pc{}}}
  \includegraphics[width=\textwidth]{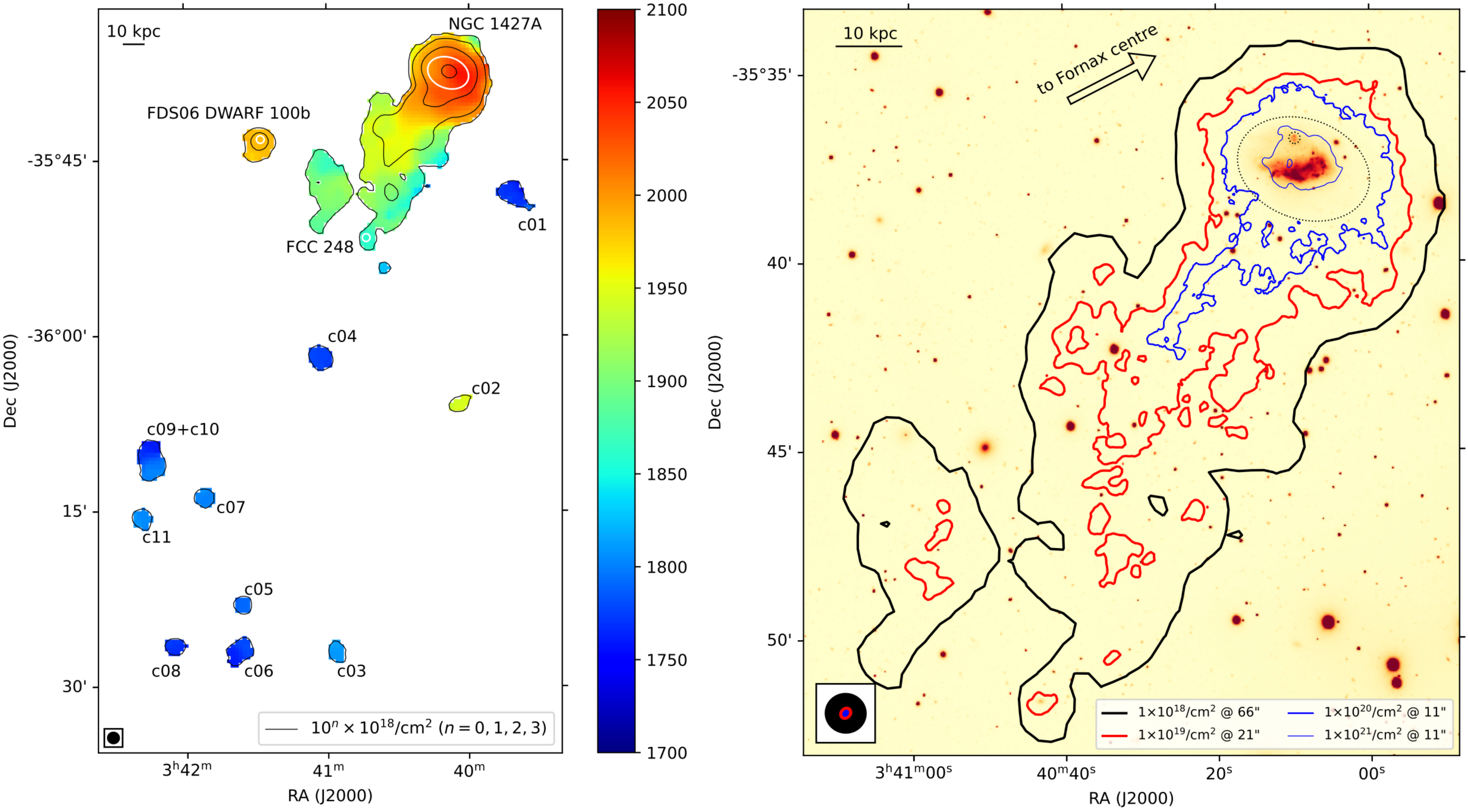}
  \centering
  \caption{Left: \HI\ velocity field of \ngc\ and its associated clouds. Centre: multi-resolution \HI\ column density map of the galactic body and tail. Right: velocity field of the \HI\ disk of \ngc~(Serra et al. 2024).}
  \label{fig:n1427a_a}
\end{figure}

\begin{figure}[t]
  \centerline{\vbox to 1pc{\hbox to 1pc{}}}
  \includegraphics[scale=0.4]{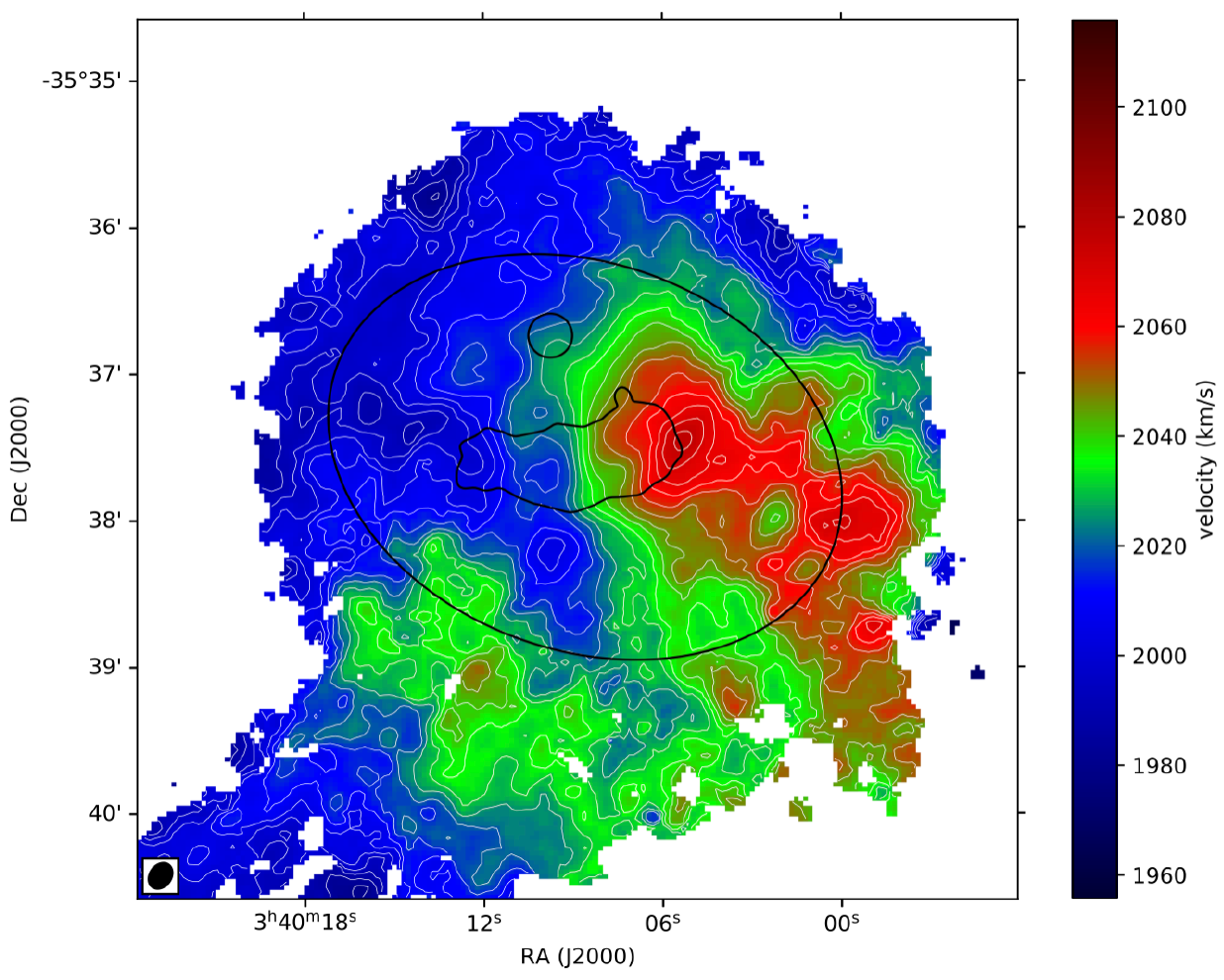}
  \centering
  \caption{Velocity field of the \HI\ disk of \ngc~(Serra et al. 2024).}
  \label{fig:n1427a_b}
\end{figure}

The detailed study of the \HI\ disk of \ngc\ and its kinematics~(Serra et al. 2024) reveals that while the galaxy is subject to the effects of ram pressure, it has also been shaped by tidal forces. The left panel of Fig.~\ref{fig:n1427a_a} shows that the extended \HI\ tail has blue-shifted kinematics. Further away, several clouds also exhibit consistent kinematics, seemingly forming the fragmented extension of the tail. The right panel of the Figure zooms in on the galaxy and its tail, showing 3-$\sigma$ \HI\ contours at different angular resolutions between 66'' and 11''. The tail has mostly low column densities ($\leq 10^{20}$~\cmsq) and an \HI\ mass of 4$\times 10^8$\msun, which represents $15\%$ of the total \HI\ mass of the galaxy. Figure~\ref{fig:n1427a_b} shows the kinematics of the \HI\ disk of \ngc, featuring a peculiar U-shape of the minor axis. 

Figures~\ref{fig:n1427a_a} and~\ref{fig:n1427a_b} provide several indications that ram-pressure is stripping the \HI\ from this galaxy. Given that, in projection, the motion of \ngc\ is strongly redshifted relative to the cluster center, a ram-pressure drag force directed toward us explains the smooth blue-shifted velocity gradient of the extended tail. In fact, ram pressure is also acting on the \HI\ embedded within the stellar body of the galaxy. The U-shape of the kinematic minor axis cannot be explained by a warp in position angle, while 3D kinematic modeling has shown that it is consistent with the same blue-shifting ram pressure acting on the tail~(Serra et al. 2024).


\begin{figure}[t]
  \centerline{\vbox to 1pc{\hbox to 1pc{}}}
  \includegraphics[width=\textwidth]{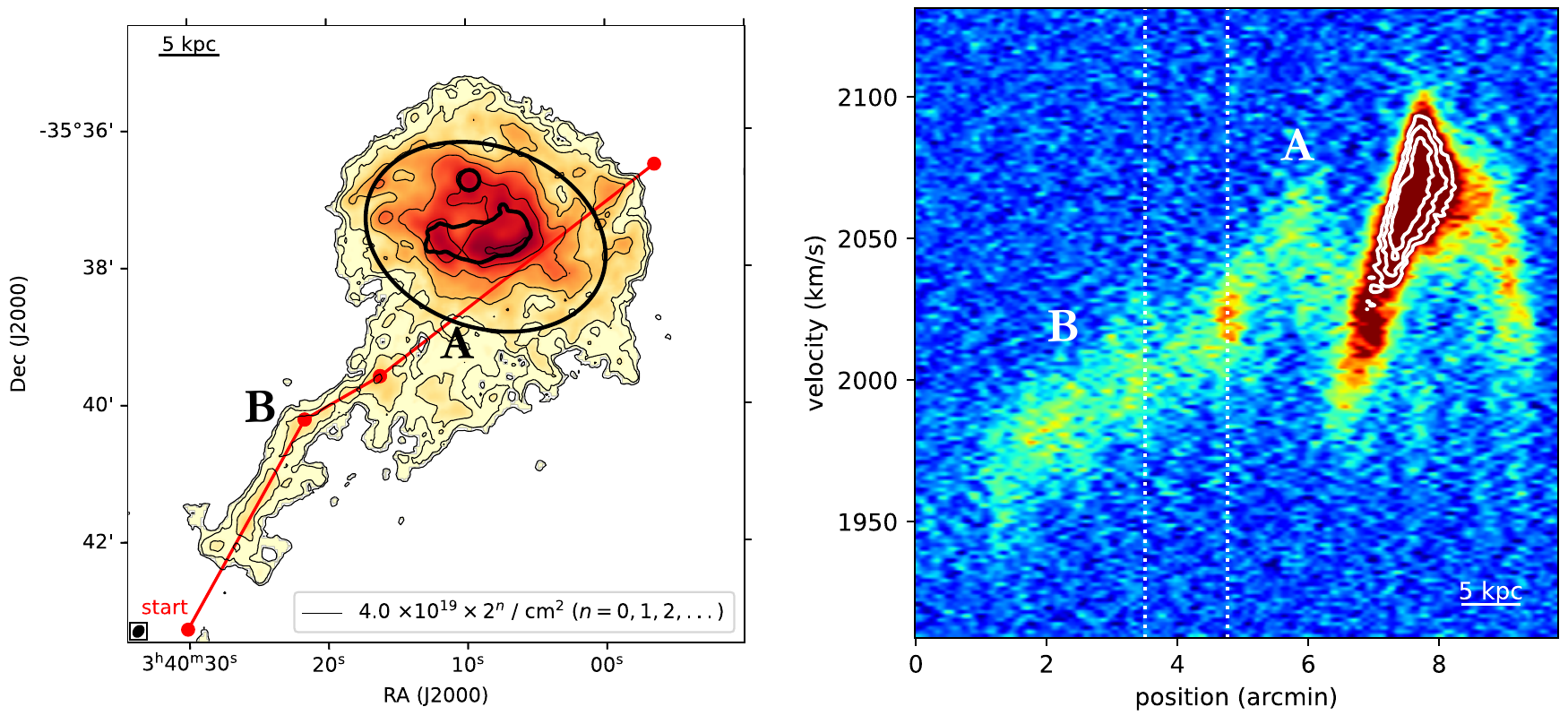}
  \centering
  \caption{Left: 10''  column density map of the tail and \HI\ disk of \ngc. Right: position-velocity diagram taken along the tail and disk of \ngc\ (along the direction marked in red in the left panel). The markers A and B show where the tail is being stripped by tidal forces and ram-pressure, respectively~(Serra et al. 2024).}
  \label{fig:pvplot}
\end{figure}

Nevertheless, other kinematic features of \ngc\ indicate tidal interactions have also affected its evolution. The clearest clear is seen in the \HI\ at the border between the stellar body and the tail. The position-velocity diagram taken along the tail and the stellar body (Fig.~\ref{fig:pvplot}) shows that the \HI\ velocity increases with radius in the initial segment of the tail (A). This is opposite to the ram pressure-driven trend along the rest of the tail (B). Therefore, in that initial segment, the \HI\ is not being stripped from the main body by ram pressure. A recent tidal interaction (possibly a merger) between two galaxies is the most plausible options. This may explain the presence of an anomalous blue stellar clump, with irregular \HI\ kinematics, in the NE edge of the disk~(Serra et al. 2024).

In conclusion, the \HI\ properties of \ngc\ indicate both tidal and hydrodynamical forces contribute in shaping the gas and stellar properties of this galaxy. Most likely, first, a merger or a high-speed collision between galaxies expelled some of the \HI\ to a large radius which then has been further stripped in the extended tail by ram pressure. 



\section{Fornax A MeerKAT observations}
\label{fornaxa}
\begin{figure}[t]
  \centerline{\vbox to 1pc{\hbox to 1pc{}}}
  \includegraphics[width=\textwidth]{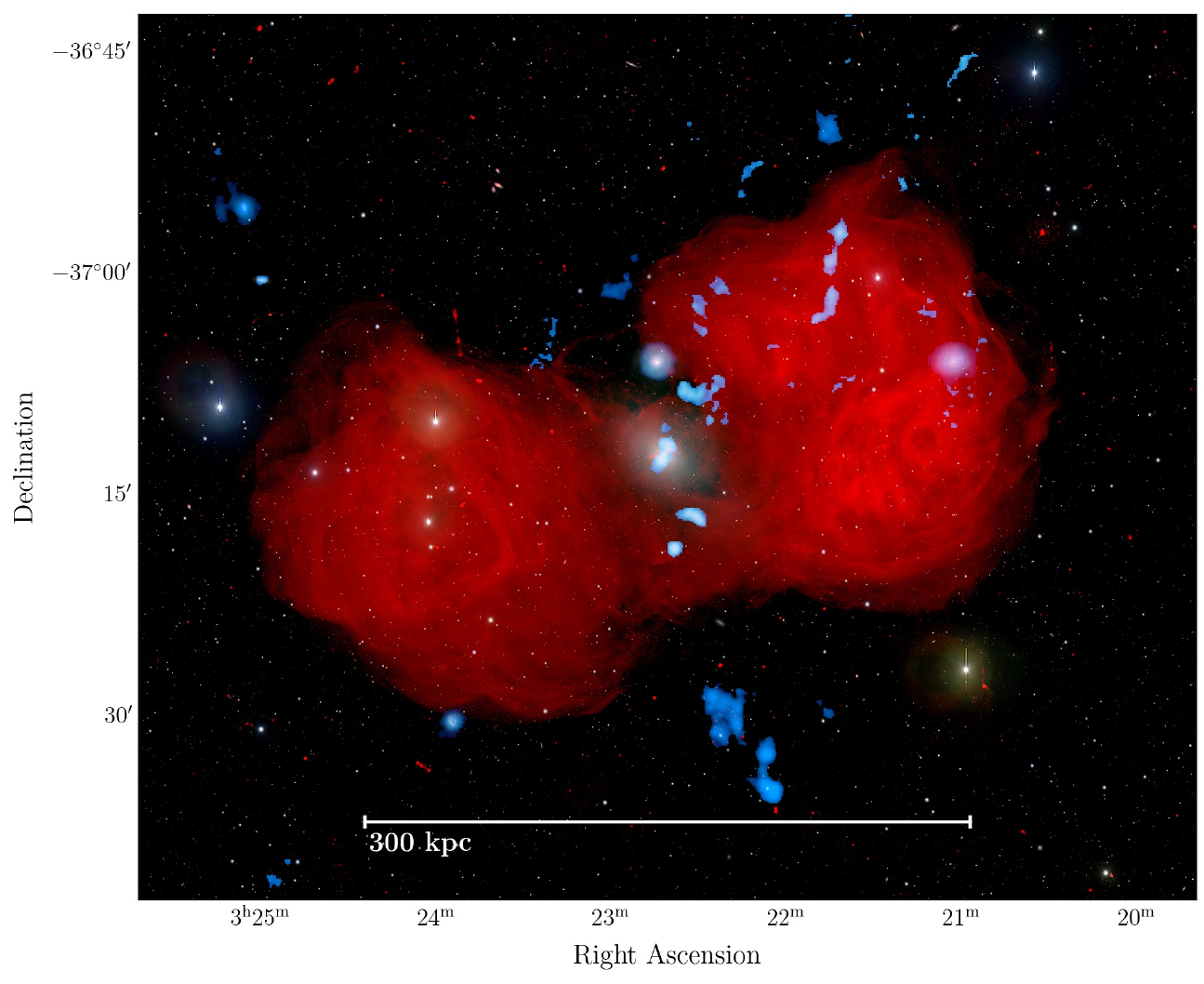}
  \centering
  \caption{MeerKAT commissioning observation of Fornax A 1.4 GHz continuum emission (red) overlaid with the neutral hydrogen clouds (blue) detected in the galaxy and its environment.}
  \label{fig:fornaxa}
\end{figure}

NGC 1316's group is currently being observed by the MFS. Due to its brightness and diffuse radio continuum emission, its AGN (Fornax A) has been an ideal test-bed for MeerKAT's performance during its commissioning. These observations delivered the deepest 1.4~GHz continuum and spectral observations of Fornax A and its group (Fig.~\ref{fig:fornaxa}). For the very first time, we discovered \HI\ associated with the radio AGN, the group members and several clouds and tails in the IGM which allowed us to understand the merger history of Fornax A~(Serra et al. 2019) and  the assembly of its group members~(Kleiner et al. 2021). In combination with other broad band continuum observations we were able to infer rapid flickering of the nuclear activity of Fornax A, revealing several events in the last few tens of Myr. The infall of multi-phase clouds from the innermost 6 kpc, traced by the \HI\, molecular and ionised gas, driven by chaotic cold accretion is regulating the fast duty-cycle of this AGN~(Maccagni et al. 2021). In combination with ASKAP polarization observations of the lobes of Fornax A, we made the first measurement of the magnetic field in an \HI\ tidal tail (visible in the western lobe) formed during a galaxy merger~(Loi et al. 2022). Further insights will come from the full depth and spectral resolution observations of the MFS.

\section{Conclusions}

First results from the MeerKAT Fornax Survey show that in the Fornax cluster galaxy tidal interactions enhance the efficiency of ram-pressure stripping, as indicated by the kinematics of the spectacular elongated \HI\ tails of the cluster members (Fig.~\ref{fig:ensemble}, Serra et al. 2023). The high resolution (10''), deep ($10^{19}$~\cmsq) observations of \ngc\ enabled a detailed analysis of the kinematics of its \HI\ disk and elongated tail~(Serra et al. 2024) demonstrating that because of the relatively low mass and high speed of the galaxy, after an encounter that first began stripping the \HI, ram pressure came into play further stripping the \HI\ off the galaxy and creating the spectacularly elongated tail. In the very near future, the Square Kilometer Array will enable the detailed studies of the physical processes driving gas removal and galaxy processing over larger statistical samples and lower mass galaxy over-densities. 

\section*{References}
\noindent Gavazzi G.\ 1989, {\it ApJ}, 346, 59.\\
Kenney J.~D.~P., van Gorkom J.~H., Vollmer B.\ 2004, {\it AJ}, 127, 3361.\\
Sancisi R., Fraternali F., Oosterloo T., van der Hulst T.\ 2008, {\it A\&ARv}, 15, 189.\\
Walter F., Carilli C., Neeleman M., {\it et al.}\ 2020, {\it ApJ}, 902, 111.\\
Jensen J.~B., Tonry J.~L., Thompson R.~I., {\it et al.}\ 2001, {\it ApJ}, 550, 503.\\
Drinkwater M.~J., Gregg M.~D., Holman B.~A., Brown M.~J.~I.\ 2001, {\it MNRAS}, 326, 1076.\\
Serra P., Maccagni F.~M., Kleiner D., {\it et al.}\ 2023, {\it A\&A}, 673, A146.\\
Mauch T., Cotton W.~D., Condon J.~J., {\it et al.}\ 2020, {\it ApJ}, 888, 61.\\
Loni A., Serra P., Kleiner D., {\it et al.}\ 2021, {\it A\&A}, 648, A31.\\
Serra P., Oosterloo T.~A., Kamphuis P., {\it et al.}\ 2024, {\it A\&A}, 690, A4.\\
Donati, J.-F., Brown, S.~F., Semel, M., {\it et. al}\ 1992, {\it A\&A}, 265, 682.\\
Iodice E., Capaccioli M., Grado A., {\it et al.}\ 2016, {\it ApJ}, 820, 42.\\
Serra P., Maccagni F.~M., Kleiner D., {\it et al.}\ 2019, {\it A\&A}, 628, A122.\\
Kleiner D., Serra P., Maccagni F.~M.,{\it et al.}\ 2021, {\it A\&A}, 648, A32.\\
Maccagni F.~M., Murgia M., {\it et al.}\ 2020, {\it A\&A}, 634, A9.\\
Maccagni F.~M., Serra P., Gaspari M., {\it et al.}\ 2021, {\it A\&A}, 656, A45.\\
Loi F., Serra P., Murgia M., {\it  et al.}, 2022\ {\it A\&A}, 660, A48.\\

\end{document}